\documentclass[11pt]{article}
\begin{document}
\title
{Velocity of sound in the relic photon sea}
\author{Miroslav  Pardy\\[0.2cm]
{\it Department of Physical Electronics}, \\
Masaryk University, Faculty of Science, Kotl\'{a}\v{r}sk\'{a} 2,\\ 611 37 Brno, Czech Republic \\
{\it e-mail: pamir@physics.muni.cz}\\
\date{\today}}
\maketitle

\vskip 5mm

\begin {abstract}
We determine the velocity of sound in the blackbody gas of photons  and in the gas of relic photons. Derivation is based on the thermodynamic theory of the photon gas and the Einstein relation between energy and mass. The spectral form for the 
n-dimensional blackbody is derived. The 
1D, 2D and 3D blackbody radiation is specified. It is mentioned  the possibility of creation of the Mach cone in case that the high energy cosmical particles moves with the speed greater than the velocity of sound in cosmical relic photon sea.

\end{abstract}

\vskip 3mm

\baselineskip 17 pt


\section{Introduction}

The relic photons form the gas of photons, or the relic photon sea.  This photon sea forms in cosmos absolute system with regard to which we can determine the absolute velocities of moving cosmical objects (bodies, stars, or cosmical rays). The velocity of Earth with regard to this system is $\approx 100 {\rm km/s}$ (Heer, 1972).
Because of the mutual elastic interaction of the relic photons, there is possibility to generate a sound in this photon gas.
 
It is well known that relic photons  discovered  in 1965 by Penzias
and Wilson forms the photon gas with the Planck distribution of temperature 2.7 of Kelvin. Sound in cosmical "vacuum" is evidently generated by collision of cosmical objects, by protuberances on the surfaces of stars, by neutron stars explosions, by Nova explosions, by collisions of galaxies, by falling asteroids on Moon, and so on.
Sound is also produce by the collision of  cosmical rays moving in the relic photon sea. It is surprising that the existence  of the cosmical sound (including the Doppler effect) was still not detected by special sensitive microphones installed in space crafts and satellites. 

Now, let us remember the Planck historical approach to the photon gas of blackbody, where blackbody can be considered as the laboratory realization of the relic photon sea.
The original Planck derivation of the blackbody radiation was based
on the relation between the entropy of the system and the internal
energy of the blackbody denoted by Planck as $U$.

While from the postulation of the relation 

$$\frac{d^2S}{dU^2} = - \frac{const}{U}\eqno(1)$$
the  Wien law follows, the a priori generalization of eq. (1) gives
new physics. The generalization of the equation (1) to be in  harmony with blackbody thermodynamics was postulated by Planck in the following form:

$$\frac{d^2S}{dU^2} = - \frac{k}{U(\varepsilon + U)},\eqno(2)$$
where $\varepsilon$ has the dimensionality of energy, $k$ is the Boltzmann constant, and formula (2)
is the approximation of the more general formula ${d^2S}/{dU^2} =
\alpha/\sum_{n}a_{n}U^{n}$ leading to exotic statistics.
 
The first integration of eq. (2) can be performed  using the integral 

$$\int\frac{dx}{x(a + bx)} = -
\frac{1}{a}\ln\left|\frac{a}{x} + b\right|.\eqno(3)$$

After integration we get the following result:

$$\frac{1}{T} =  \frac{dS}{dU} =  \frac{k}{\varepsilon }\ln
\left(\frac{\varepsilon}{U} + 1\right).\eqno(4)$$

The solution of eq. (4) is  

$$U = \frac{\varepsilon}{{\rm e}^{\varepsilon/kT} -1}.\eqno(5)$$ 

The general validity of the Wien law

$$\frac{dS}{dU} = \frac{1}{\nu}f\left(\frac{U}{\nu}\right)\eqno(6)$$
confronted with the equation (4) gives the famous Planck formula $\varepsilon = h\nu$.

The next step of Planck was to find the appropriate physical
statistical system (heuristic model) which led to the correct power spectrum of the
blackbody. This model was the thermal reservoir of the independent
electromagnetic oscillators with the discrete energies 
$\varepsilon = h\nu$. 

The Planck distribution was derived in 1900 (Planck, 1900, 1901; Sch{\"o}pf, 1978). The Planck heuristic derivation was based on the investigation of the statistics of the system of oscillators. 
Later Einstein (1917) 
derived the Planck formula from the Bohr model of atom. 
Bohr created two postulates
which define the model of atom: 1. every atom can exist in the discrete series of states in which electrons do not radiate even if they are moving at acceleration  (the postulate of the stationary states), 2. transiting electron from the stationary state to other, emits the energy according to the law $\hbar\omega = E_{m} - E_{n}$, called the Bohr formula, 
where $E_{m}$ is the energy of an electron in the
initial state, and $E_{n}$ is the energy of the final state of an
electron  to which the transition is made and $E_{m} > E_{n}$.

Einstein introduced coefficients of spontaneous  and stimulated emission $A_{mn}, B_{mn}, B_{nm}$. In case of spontaneous emission, the excited atomic state decays without external stimulus as an analog of the natural radioactivity decay. Later, quantum theory explained rigorously the process of spontaneous emission. The energy of the emitted photon
is given by the Bohr formula. In the process of the stimulated
emission the atom is induced by the external stimulus to make the
same transition. The external stimulus is a black body photon that has an energy given by the Bohr formula.

The Planck power spectral formula is as follows:

$$P(\omega)d\omega = \hbar \omega G(\omega)\frac {d\omega}{\exp{\frac {\hbar
\omega}{k_{B}T}- 1}};\quad G(\omega) = \frac {\omega^{2}}{\pi^{2}c^{3}},
\eqno(7)$$
where $\hbar\omega$ is the energy of a  blackbody photon and $G(\omega)$
is the number of electromagnetic modes inside of the blackbody, 
$k$ is the
Boltzmann constant, $c$ is the velocity of light, $T$ is the absolute temperature.

The internal density energy  of the blackbody gas is given by integration of the last equation over all  frequencies $\omega$, or

$$u = \int_{0}^{\infty}P(\omega)d\omega  = a T^{4}; \quad a = \frac{\pi^{2}k^{4}}{15\hbar^{3}c^{3}}.\eqno(8)$$

\section{The speed of sound in relic photon gas}

In order to understand the the derivation of speed of sound in gas and in the relic photon sea, we start with the derivation of the speed of sound in the real elastic rod.

Let $A$ be the cross-section of the element $Adx$ of a rod, where $dx$ is the
linear infinitesimal length on the abscissa $x$. The $\varphi(x,t)$ let be
deflection of the element $Adx$ at point $x$ at time $t$. The shift of
he element $Adx$ at point $x + dx$ is evidently

$$\varphi + \frac{\partial \varphi}{\partial x}dx. \eqno(9)$$

The relative prolongation is evidently ${\partial
  \varphi(x,t)}/{\partial x}$ .
The differential equation of motion of the rod  can be derived
by the following obligate way.  We suppose  that the force tension
$F(x, t)$ acting on
the element $Adx$ of the rod is given by the Hook law:

$$F(x,t) = EA\frac {\partial \varphi}{\partial x} ,\eqno(10)$$
where $E$ is the Young modulus of elasticity, $A$ is the cross section of the rod. We easily derive that

$$F(x+dx)-F(x) \approx EA\frac {\partial^{2} \varphi}{\partial\*x^2}\*dx \eqno(11)$$

The mass of the element $Adx$ is $\varrho Adx$, where $\varrho$ is the mass density of the rod and the dynamical equilibrium is expressed by the Newton law of force: 

$$\varrho\*Adx\varphi_{tt} = EA\varphi_{xx}dx \eqno(12) $$
or,
$$\varphi_{tt} - v^{2}\varphi_{xx}= 0, \eqno(13)$$
where

$$v = \left(\frac {E}{\varrho}\right)^{1/2} \eqno(14)$$
is the velocity of sound in the rod.

The complete solution of eq. (13) includes the initial and boundary
conditions.
We suppose that the velocity law (14) involving modulus of elasticity and mass density is valid also for gas intercalated in the rigid cylinder tube.
According to the definition of the Young modulus of elasticity 
where  $(\Delta L/L)$ is the relative prolongation of a rod, we have as an analogue for the tube of gas $\Delta V/ V$, $F \rightarrow \Delta p$, where $V$ is the volume of a gas and $p$ is pressure of a gas. Then, the modulus of elasticity is defined 
as the analogue of eq. (10). Or,

$$E = - \frac{dp}{dV}V.\eqno(15)$$

The process of the sound spreading in ideal gas is the adiabatic  thermodynamic process with no heat exchange. We use it later as a model of the sound spreading in the gas of blackbody photons. Such process is described by the thermodynamical equation

$$pV^{\kappa}= const,\eqno(16)$$
where $\kappa$ is the Poisson constant defined as $\kappa = c_{p}/c_{v}$, with $c_{p}, c_{v}$ being the specific heat under constant pressure and under constant volume. 

After differentiation of eq. (16) we get the following equation

$$dp V^{\kappa} + \kappa V^{\kappa-1} dV = 0,\eqno(17)$$
or, 

$$\frac{dp}{dV} = -\kappa \frac{p}{V}.\eqno(18)$$

After inserting of eq. (18) into eq. (15), we get from eq. (14) for the velocity of sound in gas the so called Newton-Laplace formula:

$$v = \sqrt{\kappa\frac{p}{\varrho}},\eqno(19)$$ 
where $\varrho$ is the mass density of gas.

The density of the equilibrium radiation is given by the
Stefan-Boltzmann formula

$$u = aT^{4},; \quad 
a = 7,5657.10^{-16}\frac{\rm J}{\rm K^{4}m^{3}}.\eqno(20)$$.

Then, with regard to the thermodynamic  definition of the specific heat

$$c_{v} =  \left(\frac{\partial u}{\partial T}\right)_{V} = 4aT^{3}.\eqno(21)$$ 

Similarly, with regard to the general thermodynamic theory

$$ c_{p} = c_{v} + \left[\left(\frac{\partial u}{\partial V}\right)_{T} + p\right]\left(\frac{\partial V}{\partial T}\right)_{p} = c_{v},\eqno(22)$$
because $\left(\frac{\partial V}{\partial T}\right)_{T} = 0 $ for photon gas and 
in such a way, $\kappa = 1$ for photon gas.
According to the theory of relativity, there is simple equivalence between mass and energy. Namely, $m = E/c^{2}$. At the same time, there is relation between pressure and the internal energy of the blackbody gas following from the electromagnetic theory of light $p = u/3$. So,
in our case

$$\varrho = u/c^{2} = \frac{aT^{4}}{c^{2}}; \quad p = \frac{u}{3}.\eqno(23)$$

So, after insertion of formulas in equation (23) in to eq. (19), we get the final formula for the velocity of sound in three photon sea of the blackbody is as follows:

$$ v = c \sqrt{\frac{\kappa}{3}} = \frac{c}{3}\sqrt{3},\eqno(24)$$
which is the result derived by Partovi (1993) using the QED theory applied to the photon gas. No energy signal can move with velocity greater than the speed of light. And we correctly derived  $v/c < 1$.

So, we have seen in this section, that using the classical thermodynamical model of sound in the classical gas we can easily derive some properties of the black body gas, namely the velocity of sound in it and in the relic photon sea. It is not excluded that the relic sound can be detected by the special microphones of Bell laboratories. Let us still remark
that if we use van der Waals equation of state, or, the  Kamerlingh Onnes virial equation of state, the obtained results will be modified with regard to the basic results.

\section{The n-dimensional blackbody}

The problem of the n-dimensional blackbody is related to the dimensionality of space and some ideas on the dimensionality of space  was also mentioned by author (Pardy, 2006).
The experimental facts following from QED experiments, galaxy formation and formation of the molecules DNA, prove that the external space is 3-dimensional. With regard to the Russell philosophy of mathematics, there is no possibility to prove the dimensionality of space, or, space-time, by means of  pure mathematics, because the statements of mathematics are non-existential. The existence of the external world cannot be also proved by pure mathematics. However, if there is an axiomatic system related adequately to the external world and reflecting correctly the external world, then, it is possible to do many predictions on the external world by pure logic. This is the substance of exact sciences. We know for instance that the success of special theory of relativity  is based on the adequate axiomatic system and on logic.
In case of the n-dimensional blackbody, the number of modes can be 
determined (Al-Jaber, 2003). We use here alternative and elementary derivation. In case we consider instead of  the three-dimensional blackbody the n-dimensional
blackbody, the photon energy is defined by the same manner and at the same time the statistical factor is the same as in the three-dimensional case. Only number of the electromagnetic modes $G(\omega)$ depends on
dimensionality of space. We determine in this article the Planck
blackbody law for the n-dimensional space..

The blackbody radiation is composed from the electromagnetic waves
corresponding to photons in such a way that every monochromatic wave is of the form:$ A_{\mu} = 
\varepsilon_{\mu}{\rm e}^{i{\bf k x}- i\omega t},$
where $\varepsilon_{\mu}$ is the polarization amplitude.
If we take the blackbody in the form of cube with side $L$, then it is
necessary to apply for the electromagnetic wave the boundary conditions.
It is well known that the appropriate boundary conditions are so called
periodic condition, which means for instance for x-coordinate $\exp(ik_{1}0) = \exp(ik_{1}L) = 1$,
from which follows that only specific values of $k_{1}$ correspond to the
boundary conditions, namely, $k_{1} = \frac {2\pi N_{1}}{L}; \quad N_{1} = 1, 2, 3 ... \quad.$
In case that the electromagnetic field is in a box 
of the volume  $L^{n}$, the wave vector ${\bf k}$ is
quantized and the elementary volume in the k-space is

$$\Delta_{0n} = (2\pi)^{n}/L^{n}\eqno(25)$$. 

The elementary volume of the n-dimensional k-space is evidently
the volume $d V_{n}$ between spheres with radius $k$ and $k + dk$ (Rumer et al., 1977):

$$dV_{n} = d\left(\frac {2  \pi^{n/2}}{n\Gamma\left(\frac {n}{2}\right)}
k^{n}\right) = 
\frac {2\pi^{n/2}}{\Gamma\left(\frac {n}{2}\right)}
k^{n-1}dk, \eqno(26)$$
where $\Gamma(n)$ is so called Euler gamma-function defined in the internet mathematics 
 $(http://mathworld.wolfram.com/GammaFunction.html)$ as 

$$\Gamma(x) = \int_{0}^{\infty}e^{-t}t^{x-1}dt; \quad 
\Gamma(n/2) =  \frac{(n-2)!!\sqrt{\pi}}{2^{(n-1)/2}}.\eqno(27)$$

The number of electromagnetic modes involved inside the spheres between $k$
and $k + dk$ is then, with
$\omega = ck$,  or $k = \omega/c$ and $dk = d\omega/c$,

$$G_{n}(\omega)d\omega  = 2\times \frac{dV_{n}}{\Delta_{0n}} = 2\times
\frac {1}{2^{(n-1)}}\frac {1}{\Gamma\left(\frac {n}{2}\right)}
\frac {1}{\pi^{n/2}}L^{n}\frac {\omega^{n-1}}{c^{n}} d\omega,
\eqno(28)$$
where isolated number 2 expresses the fact that light has 2
polarizations. 

For the energetic spectrum of the Planck law of the n-dimensional
black body we have

$$P_{n}(\omega) =
\hbar\omega G_{n}(\omega) \frac {1}{\exp(\frac
{\hbar\omega}{kT})-1}
= 2\times \frac {1}{2^{(n-1)}}\frac {1}{\Gamma\left(\frac {n}{2}\right)}
\frac {1}{\pi^{n/2}}\hbar \frac {\omega^{n}}{c^{n}}
\frac {1}{\exp(\frac {\hbar\omega}{kT})-1}.\eqno(29)$$

The energy density of the radiation of the n-dimensional blackbody is then

$$u_{n} = \int_{0}^{\infty}P_{n}(\omega)d\omega = A_{n}\int_{0}^{\infty}
\frac{\omega^{n}}{\exp(\frac {\hbar\omega}{kT})-1}d\omega; \quad
A_{n} = \frac {1}{2^{(n-1)}}\frac{2\hbar}{c^{n}\pi^{n/2}} \frac {1}{\Gamma\left(\frac {n}{2}\right)}.\eqno(30)$$

The integral in the last formula can be evaluated using well-known
relations (Dwight, 1961; int. 860.39)

$$\int_{0}^{\infty}\frac{x^{p}}{e^{ax}-1}dx = \frac{\Gamma(p+1)\zeta(p+1)}{a^{p+1}}
= \frac{p!\zeta(p+1)}{a^{p+1}} = \frac{p!}{a^{p+1}}\left[1 + \frac {1}{2^{p+1}} + \frac {1}{3^{p+1}} + ...\right],\eqno(31)$$
where $\zeta(p)$ is so called Riemann $\zeta$-function and $a = \hbar/kT$.

Let us test the n-dimensional Planck law and density radiation in case of n = 1, 2, and 3.

$$P_{1}(\omega) = 2\times  \frac {1}{\Gamma(1/2)}\frac {1}{\sqrt \pi}
\frac{\hbar\omega}{e^{(\frac {\hbar\omega}{kT})}-1}\frac{1}{c}\eqno(32)$$

$$P_{2}(\omega) = 2\times \frac {1}{2} \frac {1}{\Gamma(2/2)}\frac {1}{\pi}
\frac{\hbar\omega^{2}}{e^{(\frac {\hbar\omega}{kT})}-1}
\frac {1}{c^{2}} \eqno(33)$$

$$P_{3}(\omega) = 2 \times \frac {1}{4}\frac {1}{\Gamma(3/2)}\frac {1}{\pi^{3/2}}
\frac{\hbar\omega^{3}}{e^{(\frac {\hbar\omega}{kT})}-1}
\frac {1}{c^{3}}, \eqno(34)$$
and so on.

Let us remark, that  $P_{1}$ corresponds to the radiation of 1D blackbody and can be verified by long carbon nanotube at temperature $T$.  $P_{2}$ corresponds to the radiation of 2D blackbody and can be verified by the graphene sheet after some geometrical modification.  $P_{4}$  and further formulas cannot be realized in the 3D space with the adequate blackbody.

$$u_{1} = A_{1}\int_{0}^{\infty}\frac{x}{e^{ax}-1}dx = A_{1}\left(\frac{kT}{\hbar}\right)^{2} 1! \zeta(2) =
A_{1}\left(\frac{kT}{\hbar}\right)^{2}\frac {\pi^{2}}{6}; \quad  A_{1} = \frac{2\hbar}{c\pi^{1/2}} \frac {1}{\Gamma\left(\frac {1}{2}\right)}\eqno(35)$$

$$u_{2} = A_{2}\int_{0}^{\infty}\frac{x^{2}}{e^{ax}-1}dx =
 A_{2}\left(\frac{kT}{\hbar}\right)^{3} 2! \zeta(3) = A_{2}\left(\frac{kT}{\hbar}\right)^{3} 2 \times 1,202;\quad  A_{2}  = \frac{\hbar}{c^{2}\pi} \frac {1}{\Gamma\left(1\right)}\eqno(36)$$

$$u_{3} = A_{3}\int_{0}^{\infty}\frac{x^{3}}{e^{ax}-1}dx = 
A_{3}\left(\frac{kT}{\hbar}\right)^{4} 3!\zeta(4) =
A_{3}\left(\frac{kT}{\hbar}\right)^{4}6\frac {\pi^{4}}{90}; A_{3} =
 \frac{\hbar}{2c^{3}\pi^{3/2}}
 \frac {1}{\Gamma\left(\frac {3}{2}\right)}\eqno(37)$$
and so on, where we used tables of Dwight (1961) with formulas 48.002, 48.003, 48.004
for $\zeta(2) = \pi^{2}/6, \zeta(3) = 1,2020569032, \zeta(4) = \pi^{4}/90$

Let us remark that the formula (37) is identical with formula (8) with regard
to relation $\Gamma(x +1) = x\Gamma(x)$, or, $\Gamma(3/2) = \Gamma(1/2 + 1) = (1/2)\Gamma(1/2)  = (1/2)\pi^{1/2}$, and it  is the proof of the correctness of derived formula $u_{3}$.  

\section{Discussion}

Our derivation of the light velocity in the relic photon sea was based on the classical thermodynamical model with the adiabatic process ($\delta Q = 0$), controlling the spreading of sound in the relic sea. The problem was not solved by Einstein, because only QED, elaborated many years after Einstein article, gives information on the photon-photon interaction and on the discovery of the gas of relic photons. In other words, Einstein was not motivated for such activity. Partovi (1993) derived additional radiation corrections to the Planck distribution formula and the additional correction to the speed of sound in the relic photon sea. His formula is of the form:

$$v_{sound} = \left[1 - \frac{88\pi^{2}\alpha^{2}}{2025}\left(\frac{T}{T_{e}}
\right)^{4}\right]\frac{c}{\sqrt{3}}, \eqno(38)$$
where $\alpha$ is the fine structure constant and $T_{e} = 5.9$ G
Kelvin. We see that our formula is the first approximation in the Partovi expression.

There is rigorous statistical theory of transport of sound energy in gas based
on the Boltzmann equation (Uhlenbeck and Ford, 1963).  After
application of Boltzmann equation to the photon gas, or, relic photon
gas we can expect the rigorous results with regard to fact that the cross-section of the photon-photon interaction is very small. Namely (Berestetzkii et al., 1999):

$$\sigma_{\gamma \gamma} = 4,7 \alpha^{4}\left(\frac{c}{\omega}\right)^{2};\quad
\hbar\omega\ll mc^{2},\eqno(39)$$
and

$$\sigma_{\gamma \gamma} = \frac{973}{10125 \pi}
\alpha^{2}r_{e}^{2}\left(\frac{\hbar\omega}{mc^{2}}\right)^{6};\quad
\hbar\omega\gg mc^{2},
\eqno(40)$$
where $r_{e} = e^{2}/mc^{2} = 2,818 \times 10^{-13}$ cm is the classical radius
of electron and $\alpha = e^{2}/\hbar c$ is the fine structure constant with numerical value $1/\alpha = 137,04$.

No doubt, the solution of the
Boltzmann equation gives not only the existence of sound waves in the statistical system of particles, however the sound waves in in interstellar space.

The cosmical rays including relic photons were predicted 
by Gamow as a consequence of the Big Bang. It is evident that  the
Mach cone is created in case that the high energy cosmical particles move with the speed greater than the velocity of sound in cosmical relic photon sea.

 While there is heuristic models by Einstein and Planck  for the derivation of the Planck Law, the similar  derivation of cosmical  relic photons was not performed. It is not excluded that relic photon are created during cosmological expansion and acceleration by the Hawking effect. 
   
By the early 1970' it became clear that the CMB sky is of the dipole
form (it is hotter in one
direction and cooler in the opposite direction with the temperature
difference being a few mili Kelvin). The dipole form can be explained
by the motion our galaxy with regard to the rest of universe. However,
at some level one expects to see irregularities, or anisotropies of
CMB, and then it is not excluded that the generalized Planck formula
will be appropriate for the description of CMB. If we respect such
anisotropy, then inevitable result will be that there is also
anisotropy of sound in relic photon sea.

\vspace{5 mm}

REFERENCES

\vspace{5 mm}

\noindent
Al-Jaber, Sami M. (2003). Planck's spectral distribution law in N dimensions, Int. Journal
of Theor. Phys. {\bf 42}, No. 1, 111.\\
Berestetzkii, V. B., Lifshitz, E. M. and Pitaevskii, L. P. (1999). 
{\it Quantum electrodynamics},
(Butterworth-Heinemann, Oxford).\\
Dwight, H. B. (1961). {\it Tables of integrals}, (New York, The
Macmilan Company).\\
Einstein, A. (1917). Zur quantentheorie der Strahlung, Physikalische
Zeitschrift, {\bf 18}, 121.\\
Heer, C. V. (1972). {\it Statistical mechanics, kinetic theory and
  stochastic processes}, 
(Academic press, New York).\\
http://mathworld.wolfram.com/GammaFunction.html\\
Pardy, M. (1906). Radiation of the blackbody in the external
field,\\ arXiv:quant-ph/0509081v2 6 Mar 2006.\\
Partovi, M. Hossein (1994).
QED corrections to Planck's radiation law and photon thermodynamics,
Phys. Rev. D {\bf 50}, No. 2,  1118-1124.\\
Planck, M. (1900). Zur Theorie des Gesetzes der Energieverteilung im Normalspektrum,
Verhandlungen deutsch phys., Ges., {\bf 2}, 237.; ibid:
(1901). Ann. Phys., {\bf 4}, 553.\\
Rumer, Yu. B. and Ryvkin, M. Sch. (1977). {\it Thermodynamics, statistical physics, kinetics}, (Nauka, Moscow).\\
Sch{\"o}pf, H-G. (1978).{\it Theorie der W{\"a}rmestrahlung in
 historisch-kritischer Darstellung},\\(Alademie/Verlag, Berlin).\\
Uhlenbeck, G. E. and Ford, G. W. (1963). {\it Lectures in statistical
physics}, (American mathematical society, Providence, Rhode Island).

\end{document}